\documentclass{article}

\usepackage{ml4cps}

\usepackage[utf8]{inputenc} % allow utf-8 input
\usepackage[T1]{fontenc}    % use 8-bit T1 fonts
\usepackage{hyperref}       % hyperlinks
\usepackage{url}            % simple URL typesetting
\usepackage{booktabs}       % professional-quality tables
\usepackage{amsfonts}       % blackboard math symbols
\usepackage{nicefrac}       % compact symbols for 1/2, etc.
\usepackage{microtype}  
\usepackage{amsmath} 
\usepackage{cleveref}       % smart cross-referencing
\usepackage{lipsum}         % Can be removed after putting your text content
\usepackage{graphicx}
\usepackage[numbers]{natbib}
\usepackage{doi}
\usepackage{algorithm}
\usepackage{algpseudocode}
\usepackage{amsmath}

\title{Steganographic Embeddings As an Effective Data Augmentation}

\date{}

\newif\ifuniqueAffiliation
\uniqueAffiliationtrue

\ifuniqueAffiliation % Standard variant of author block
\author{ Nicholas DiSalvo\\
	\texttt{nickd16718@gmail.com} \\
}
\else
\usepackage{authblk}

\affil[1]{Affiliation, Address}
\affil[2]{Affiliation, Address}
\fi

\renewcommand{\shorttitle}{Steganographic Embeddings As an Effective Data Augmentation}

\hypersetup{
pdftitle={Steganographic Embeddings As an Effective Data Augmentation},
pdfsubject={Steganpgrahy applied to Image Classification},
pdfauthor={Nicholas DiSalvo},
pdfkeywords={Image Steganography Least Significant Bit (LSB) Data Augmentation Deep Neural Networks Image Classification CIFAR-10 Tiny-ImageNet STL-10 Training Efficiency Computer Vision},
}

\begin{document}

\maketitle

\begin{abstract}
Image Steganography is a cryptographic technique that embeds secret information into an image, ensuring the hidden data remains undetectable to the human eye while preserving the image's original visual integrity. Least Significant Bit (LSB) Steganography achieves this by replacing the k least significant bits of an image with the k most significant bits of a secret image, maintaining the appearance of the original image while simultaneously encoding the essential elements of the hidden data. In this work, we shift away from conventional applications of steganography in deep learning and explore its potential from a new angle. We present experimental results on CIFAR-10 showing that LSB Steganography, when used as a data augmentation strategy for downstream computer vision tasks such as image classification, can significantly improve the training efficiency of deep neural networks. It can also act as an implicit, uniformly discretized piecewise linear approximation of color augmentations such as (brightness, contrast, hue, and saturation), without introducing additional training overhead through a new joint image training regime that disregards the need for tuning sensitive augmentation hyperparameters.
\end{abstract}

% add keywords

\begin{figure}[h!]
    \centering
    \includegraphics[width=1\linewidth]{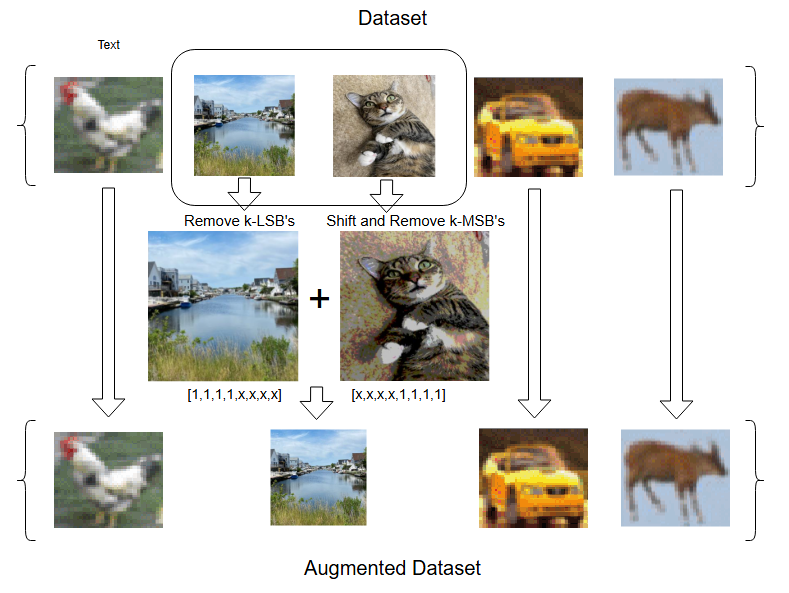}
    \caption{This is a visual demonstration of the LSB Steganography Augmentation Strategy. When creating batches for our dataset, we at random take image pairs from the dataset and one inside the other to create an embedded steganographic image we will use to train our model.}
    \label{fig:figure1}
\end{figure}

\section{Introduction}

Deep Neural Networks are prone to overfitting on their training datasets, leading to reduced generalization to unseen examples outside the dataset on which they are trained. This issue has persisted from the inception of deep learning to modern LLMs \cite{32}. Data augmentation is a technique in deep learning used to reduce overfitting by increasing the size and diversity of training data through transformations applied to the original data \cite{7}. Examples of simple augmentations include image flipping, rotation, and scaling. These augmentations provide the model with variations of the original data, increasing the dataset size and enabling the model to learn invariance to geometric transformations \cite{2}. More advanced augmentations include mixing and erasing techniques, which create new samples by blending images or removing random parts of an image, sometimes filling the gaps with selected areas from other images \cite{4}. Additionally, deep generative models (e.g., GANs, Variational Autoencoders (VAEs), Diffusion Models, Neural Radiance Fields (NeRFs), and Flow-Based Models) can also be considered data augmentation methods because they generate synthetic data that follow the original training data distribution, providing rich new samples for various downstream vision tasks \cite{5}.

Image Steganography is a cryptographic technique that embeds secret information into an image, ensuring the hidden data remains undetectable to the human eye while preserving the image's original visual integrity. Traditionally, steganography has been used for secure communication and data protection, enabling information to be transmitted covertly \cite{22}. The classic method of steganography is Least Significant Bit (LSB) Steganography, which hides images within other images by replacing the k least significant bits of an image with the k most significant bits of a secret image \cite{7}. Other techniques for embedding images include discrete cosine transforms (DCT), pixel value differencing (PVD), palette-based steganography, histogram shifting, and more \cite{8}.

Many studies on steganography in the context of deep learning have focused on enhancing naive methods with deep neural networks. These approaches often utilize encoder-decoder systems, such as Variational Autoencoders (VAEs), to encode images into a compact latent space, with the decoder subsequently reconstructing the images \cite{9}. Some methods employ GANs, which leverage a generator and discriminator to produce high-quality images with neural embeddings \cite{10}. In this work, we take a different approach by using naive Least Significant Bit (LSB) steganography as a form of data augmentation rather than attempting to improve traditional steganographic techniques. This augmentation not only expands the dataset but also enhances training efficiency. By embedding one image into another using LSB steganography, we generate new data points that retain the same number of bits as a typical training image while containing a near-perfect replica of one image alongside a degraded version of the other. Additionally, this approach serves as an implicit, quantized piecewise linear approximation of classic color-jitter augmentations such as saturation, contrast, brightness, and hue adjustments, by introducing implicit degradation through the nature of steganographic images. This helps improve model robustness while ensuring that the overall training time remains unchanged compared to training without explicit augmentations \cite{11}.

\section{Related Works}
Classic data augmentations, such as rotations, translations, flipping, cropping, hue and saturation changes, etc., for training deep learning networks have been explored by \cite{7}, \cite{9}, \cite{24}, \cite{26}, \cite{27}. More advanced data augmentation strategies, such as mixup, cutmix, and cutout, have been explored by \cite{24}, \cite{25}, \cite{26}, \cite{30}. These strategies aim to combine various images from the training dataset to create augmented images that enhance the network's diversity. These strategies are similar to our approach in that regard, but they differ in the methodology used to achieve these combinations \cite{9}, \cite{6}, \cite{11}.

Steganography has also been extensively explored in the context of deep learning. The works in \cite{12}, \cite{13}, \cite{15}, \cite{16} have used neural techniques to improve the concealment of data in the naive implementation of steganography. These works \cite{18}, \cite{17}, \cite{22} employ generative models such as GANs, encoder-decoder architectures like the Variational Autoencoder (VAE), etc., to create diverse samples from the training distribution \cite{20}, \cite{21}, \cite{14}. These ideas have been explored together to achieve data augmentation, but to our knowledge, no works have adapted naive steganography as an embedding augmentation for downstream computer vision tasks, such as classification, and have shown their resemblance to traditional augmentation strategies in this context.

\section{Steganography as an Augmentation} 
This section will describe the approach of using LSB Steganography as a powerful data augmentation strategy for training deep neural networks for computer vision tasks. 
\subsection{The Algorithm}
To formalize the proposed strategy, when processing a batch of images for input into our neural network, we will, with a probability p, select a pair of images instead of a single one. Subsequently, we will embed one image into the other using Least Significant Bit (LSB) steganography with k bits. This process entails replacing the k least significant bits of the first image with the 
k most significant bits of the second image, resulting in a composite image. This composite image will then be incorporated into our batch. Consequently, approximately D * p of the images within our input data batch will contain the embedded information of two images derived through steganography. Refer to Figure 1 for an illustrative representation.

In this study, we apply this augmentation technique within the domain of image classification. To accommodate the presence of multiple classes within a single image, we modify our labels to adopt the form of one-hot vectors. Specifically, we assign a value of 1 to each class represented in the image, thereby enabling simultaneous accounting of multiple classes in our dataset.

\subsection{Efficiency of Steganography as an Augmentation}
When preparing image data for neural network training, images are typically represented as 32-bit or 16-bit floating-point tensors, though quantization to 8-bit integer tensors is common to reduce memory and computational overhead. Regardless of the chosen bit depth, each tensor element conventionally encodes a single image with a fixed numerical precision, limiting the information density per data point. The primary motivation for employing steganography as a data augmentation technique lies in maximizing the informational utility of these bits during training. Traditional augmentations, such as flips, rotations, or color jitter, enhance model generalization and test-time performance by expanding the dataset through additional transformed samples, thereby increasing computational cost proportional to the augmented dataset size. In contrast, steganography embeds a pair of images, a primary (cover) image and a secondary (hidden) image, within a single data point, preserving the original dataset size while enriching the gradient signal available during backpropagation. Despite each training sample containing two images, the input dimensionality remains fixed, enabling the model to simultaneously extract features from both images without inflating the number of training samples or iterations. This approach optimizes bit usage per data point, enhancing feature learning and training efficiency without extending the training duration. In the following sections, we fix the number of epochs and rigorously evaluate the impact of this steganography-based augmentation on training efficiency, demonstrating its efficacy as a computationally efficient alternative to traditional methods.

\subsection{Steganography as a Discrete Linear Approximator of Color Augmentations}
\begin{figure}[h!]
    \centering
    \includegraphics[width=1\linewidth]{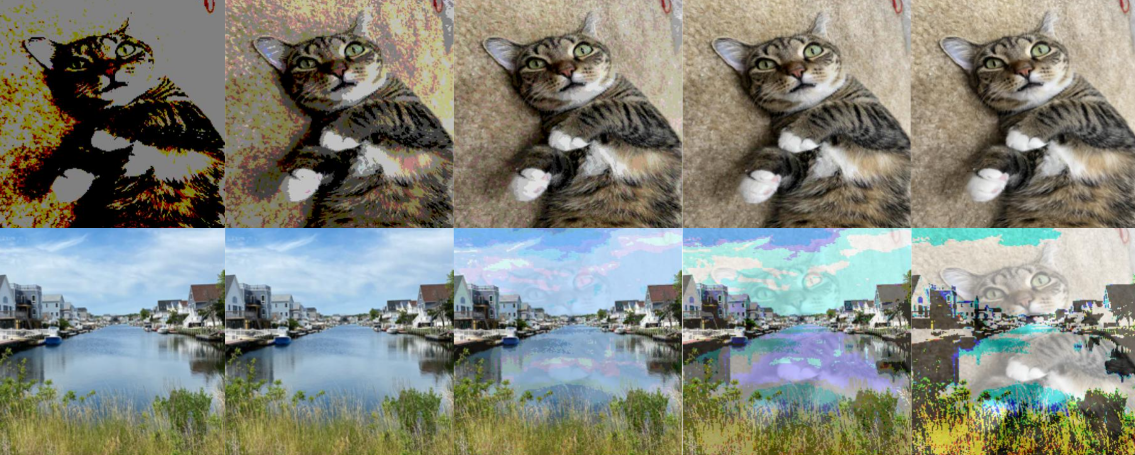}
    \caption{Number of Bits For Steganography: [1,2,3,5,7]\\From this visual, we see 
    stegangraphic embeddings implicitly add color augmentations such as brightness, contrast, saturation, and hue to our model with no added overhead.}
    \label{fig:figure2}
\end{figure}

In addition to the added efficiency provided by steganography as an augmentation, we also gain additional color-altering augmentations at no extra cost. This is because, depending on the number of bits we use, the embedded or cover image in which we are embedding an image can change in brightness, contrast, saturation, and hue based on the different pixel values resulting from removing or adding bits to the image. We, therefore, introduce color augmentations to our dataset (see \hyperref[fig:figure2]{Figure 2}) without the overhead of expanding the training time of our model by passing the images separately with color alterations applied. 

To understand the mathematical equivalance of these augmentations, let us take a look at the underlying pixel to pixel mappings in the RGB space of these augmentations. \\
**Note, more robust methods of calculating contrast and saturation take place in the HSV/HSL color spaces rather than RGB, but for our purpose, the general RGB intuitive mappings will be enough to demonstrate the significance. 

Consider a pixel intensity 
\(I \in [0, 255]\)
in a given RGB channel. Common color augmentations can be expressed as follows:

Brightness: Adjusts intensity by a constant bias 
\(b\),
\[
I' = \text{clamp}(I + b, 0, 255),
\]
Contrast: Scales intensity around the midpoint (128) by a factor 
\(s > 0\),
\[
I' = \text{clamp}(128 + s (I - 128), 0, 255).
\]
Saturation: Adjusts intensity relative to the grayscale value 
\(G = 0.299R + 0.587G + 0.114B\)
with a factor 
\(c > 0\),
\[
I' = \text{clamp}(G + c (I - G), 0, 255).
\]
These transformations can be generalized as a linear function of the form:
\[
I' = \alpha * I + \beta,
\]
Where, 
\(\alpha\)
controls scaling (contrast/saturation), and 
\(\beta\)
introduces a shift (brightness bias). Since 
\(I\), 
\(\alpha\), and 
\(\beta\)
are real-valued, 
\(I'\)
is continuous over 
\([0, 255]\)
before clamping, allowing fine-grained adjustments but requiring careful tuning of 
\(\alpha\)
and 
\(\beta\).
In LSB steganography, we embed data by replacing the 
\(k\)
least significant bits of a pixel value 
\(I\)
with bits from the hidden image. Mathematically, this is a quantization process:
\[
I' = \left\lfloor \frac{I}{2^k} \right\rfloor \cdot 2^k,
\]
where 
\(\left\lfloor x \right\rfloor\)
denotes the floor function. The resulting 
\(I'\)
takes values in the discrete set:
\[
I' \in \{0, 2^k, 2 \cdot 2^k, \ldots, \left\lfloor \frac{256}{2^k} \right\rfloor \cdot 2^k\}.
\]
This partitions the intensity range 
\([0, 255]\)
into 
\(N = \left\lfloor \frac{256}{2^k} \right\rfloor\)
bins of width 
\(2^k\),
with each 
\(I\)
mapped to the nearest lower multiple of 
\(2^k\).
The probability of 
\(I'\)
taking any specific value in this set, under a uniform input distribution for 
\(I\),
is:
\[
p(I') = \frac{1}{N} \approx \frac{2^k}{256},
\]
indicating a uniform discrete distribution over the quantized levels.

\begin{figure}
    \centering
    \includegraphics[width=1\linewidth]{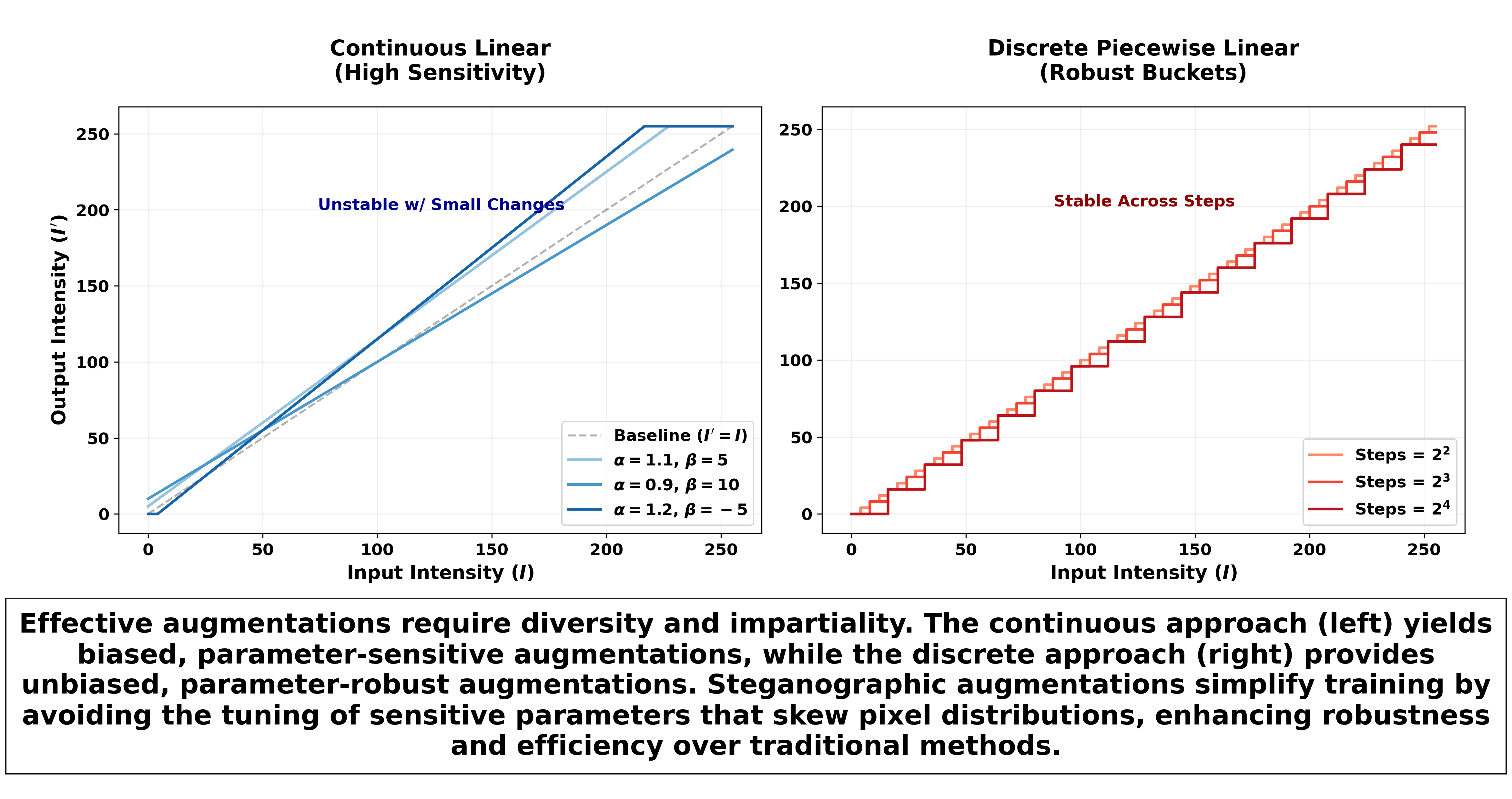}
    \label{fig:figure3}
\end{figure}

Now that we understand how I' is computed, lets analyze $\Delta I$, the change in I from steganographic augmentations. Note, $\Delta I$ in the continuous case was the result of a sensitive continous linear function over the interval [0,255]. Now, we see that 
$$\Delta I = I - I' = I  - (\left\lfloor \frac{I}{2^k} \right\rfloor \cdot 2^k) = I - (I \text{ mod } 2^k)$$

From these equations and \hyperref[fig:figure3]{Figure 3}, we see that LSB steganography naturally approximates traditional color augmentations by discretizing the intensity space into \(\frac{256}{2^k}\) evenly spaced bins of size \(2^k\) through uniform quantization, as defined by the set \(I' \in \{0, 2^k, 2 \cdot 2^k, \ldots, \left\lfloor \frac{256}{2^k} \right\rfloor \cdot 2^k\}\). Unlike continuous augmentations, which scale \(I\) with \(\alpha\) and shift it with \(\beta\) (e.g., \(I' = \alpha I + \beta\)) but may bias toward midtones or grayscale and require careful tuning of hyperparameters, LSB steganography applies a discrete perturbation, \(-(I \mod 2^k)\), acting as a brightness-like shift without scaling. This results in a uniform distribution of \(I'\), ensuring no preference for specific intensity ranges and eliminating bias inherent in traditional methods. By varying \(k\) randomly during training, we introduce diversity akin to unbiased sampling \(\alpha\) and \(\beta\) uniformly, bypassing the need for manual hyperparameter adjustments that often affect model generalization. This purely quantized approach thus provides a robust, efficient, and principled alternative for data augmentation, maintaining consistency and impartiality across the training process.

\subsection{Additional Benefits of Steganographic Augmentations}
Another compelling property of this augmentation is that it encourages the model to distribute its attention more uniformly across all bits of an image. In a conventional training setup, it is natural to assume that a model primarily focuses on the most significant bits, as they encode the dominant perceptual information, precisely the information human labelers rely on when annotating datasets. Since the higher-order bits control the overall structure and appearance of an image, they inherently carry more weight in determining its class. However, by embedding additional information in the least significant bits, we introduce an implicit regularization effect, compelling the model to extract meaningful features from all available bits rather than disproportionately weighting the most significant ones.

This concept aligns closely with the intuition behind dropout \cite{33}, where randomly dropping units during training prevents the model from over-relying on specific neurons, thereby promoting robustness and more distributed feature learning. Similarly, in our augmentation strategy, the model cannot afford to ignore lower-bit information, as doing so would lead to suboptimal performance on the embedded secondary image. In this sense, our approach acts as a form of bit-level dropout, ensuring that every bit contributes meaningfully to the learning process. By forcing the network to process information across the entire bit spectrum, we encourage a more resilient and generalizable feature representation, mitigating the risk of overfitting to dominant patterns found in high-order bits alone.

\section{Experimental Details}
This section describes the experimental details of the proposed steganography embedding augmentation strategy. We apply our data augmentation strategy in the context of image classification on CIFAR-10 and test it on a small, approximately 2-million-parameter CNN model by Keller Jordan in his recent study on CIFAR-10 training speedruns (Airbench).

\subsection{Dataset, Loss Function, and Model Architectures}
For our data labels, since steganographic images may be present in our input training batches, we adjust our integer labels to one-hot vectors to account for multiple classes within a single image. Consequently, we optimize the likelihood of our dataset by minimizing binary cross-entropy as our loss function, effectively handling the output vector of raw logits for each label. We select CIFAR-10 as our dataset due to its simplicity and manageable size, enabling the training of thousands of runs on minimal hardware. For our model architecture, we adopt the naive implementation of Keller Jordan's Airbench model, which has been used in recent CIFAR-10 training speedruns \cite{35}.

\subsection{Optimization}
One important note for our study is that we do not aim to improve the benchmarks of image classification; our goal is simply to demonstrate the power of our augmentation strategy. For this reason, we choose a simple yet diverse optimizer and learning rate. We use AdamW with betas=(0.9, 0.95), weight decay = 0.01, and a learning rate of 2e-4. All of our experiments are conducted in PyTorch Lightning using a single Nvidia RTX 4060 Ti, and all experiments can be reproduced from this \href{https://github.com/nickd16/steganographic-augmentations}{repository}. To back up our experimental evidence, we trained and averaged our experiments from over 10000 full training rounds. 

\subsection{Steganography Hyperparameters}
For our experiments, we explored various bit depths for image steganography and empirically determined that keeping them uniformly random during training yielded the best results. For 8 bit images, we chose k uniformly from \{1,2,3,...,7\}. In addition to bit depth, we selected the probability of applying our augmentation, which we set to 0.5. Setting this probability much higher would cause the model to rely too heavily on steganographic images, potentially impairing its performance during testing and inference. Our primary objective in applying this augmentation is to enhance training efficiency while improving generalization. However, real-world downstream tasks still involve single images, making it essential for the model to also encounter normal images during training to prevent overfitting.

\begin{figure}
    \centering
    \includegraphics[width=1\linewidth]{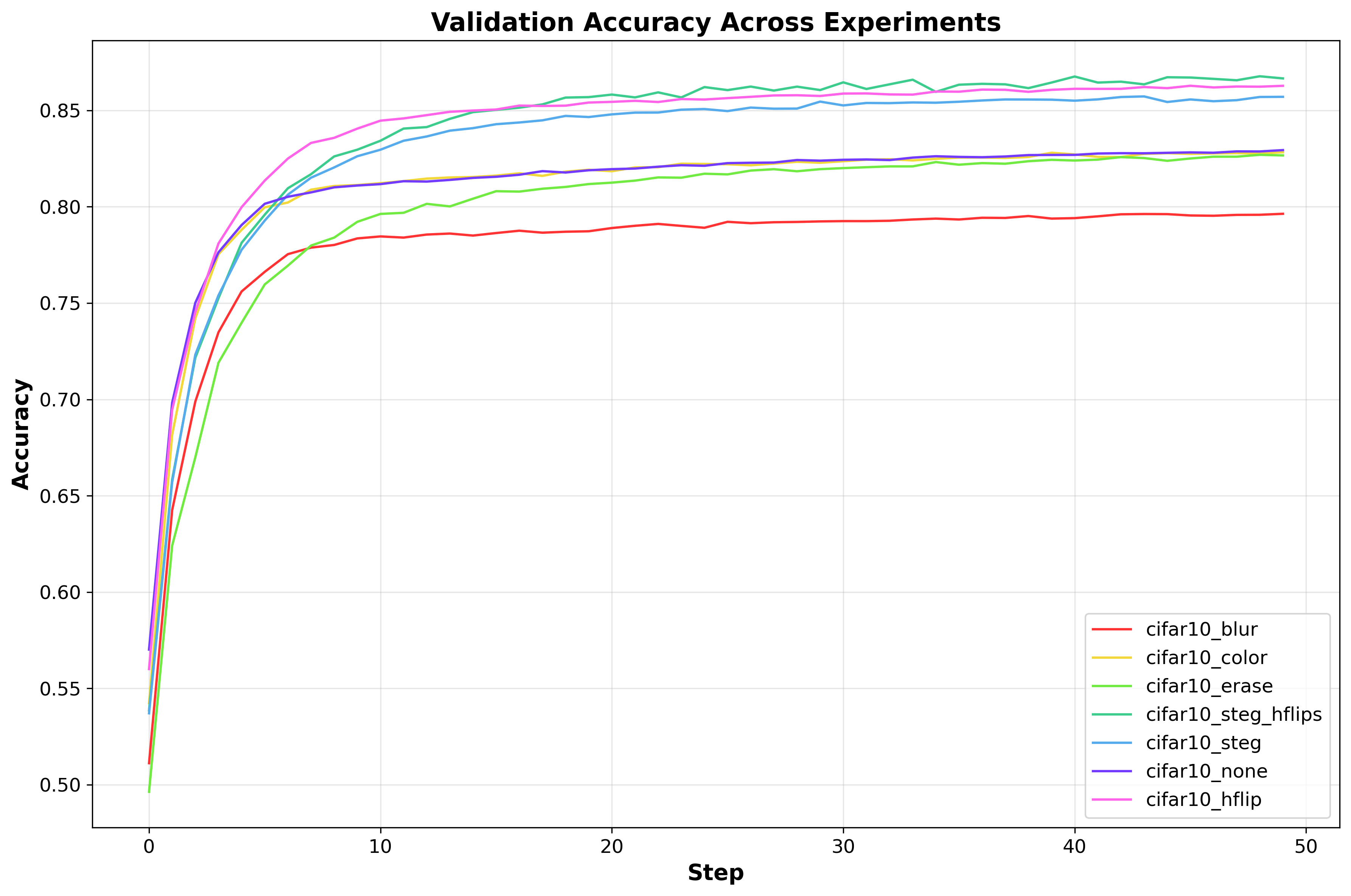}
    \caption{Results comparing steganographic augmentations to color-jitter (saturation, brightness, contrast, hue), Gaussian blur, random erasing, horizontal flips, no augmentations, and a combined horizontal flip + steganography}
    \label{fig:figure4}
\end{figure}

\subsection{Results}
The results of our experiments can be found at \hyperref[fig:figure4]{Figure 4}) comparing steganographic augmentations to other augmentation techniques, including color jitter (saturation, brightness, contrast, hue), Gaussian blur, random erasing, horizontal flips, no augmentations, and a combined horizontal flip + steganography. As shown in the graph, which presents an average over 10,000 full training iterations, steganography as an augmentation outperforms all others except horizontal flips, which have been empirically the most effective out-of-the-box augmentation. Since our findings demonstrate that steganographic augmentations approximate a continuous, pixel-geometry-preserving linear distribution, it is reasonable to assume that combining them with a geometric transformation, such as a horizontal flip, yields the best overall performance ("getting the best of both worlds").

\section{Conclusion}
In this paper, we demonstrate that our novel LSB steganographic augmentations enhance model efficiency during training. By ensuring equal representation of all bits in the input data, this approach mitigates overfitting while simultaneously providing implicit color augmentations at no additional cost. Additionally, it enables the extraction of gradient signals from multiple images concurrently. We validate this method through image classification on CIFAR-10 using a basic 2-million-parameter CNN model. Furthermore, we show that these augmentations function as an implicit, uniformly discretized piecewise linear approximation of color transformations, such as brightness, contrast, hue, and saturation, without introducing additional training overhead, facilitated by a new joint image training regime. In future work, we aim to assess whether these augmentations improve performance on specific downstream computer vision tasks. We also plan to investigate the effects of using the paradigm of utilizing bits into other modalities. 

\section*{Acknowledgments}
All work in this paper was done as pure individual research without any institutional backing. Special thanks to my pet cat "Boots" for the beautiful visuals! :) 

\bibliographystyle{abbrv}
%\bibliography{references}  %%% Uncomment this line and comment out the ``thebibliography'' section below to use the external .bib file (using bibtex) .

%%% Uncomment this section and comment out the \bibliography{references} line above to use inline references.

\end{document}